\documentclass[prd,floatfix,preprintnumbers,showpacs]{revtex4}
\usepackage{graphicx}
\usepackage{dcolumn}
\usepackage{bm}
\usepackage{hyperref}
\usepackage{subfigure}

\def\lsim{\mathrel{\mathop
  {\hbox{\lower0.5ex\hbox{$\sim$}\kern-0.8em\lower-0.7ex\hbox{$<$}}}}}
\def\gsim{\mathrel{\mathop
  {\hbox{\lower0.5ex\hbox{$\sim$}\kern-0.8em\lower-0.7ex\hbox{$>$}}}}}

\begin{document}
\newcommand{\beq}{\begin{equation}}
\newcommand{\eeq}{\end{equation}}
\newcommand{\beqra}{\begin{eqnarray}}
\newcommand{\eeqra}{\end{eqnarray}}
\newcommand{\Ocdm}{\Omega_{\rm cdm}}
\newcommand{\ocdm}{\omega_{\rm cdm}}
\def\ds{\displaystyle}
\def\ts{\textstyle}
\def\ss{\scriptstyle}
\def\sss{\scriptscriptstyle}
\newcommand{\Ob}{\Omega_{\rm b}}
\newcommand{\ob}{\omega_{\rm b}}
\newcommand{\Om}{\Omega_{\rm m}}
\newcommand{\om}{\omega_{\rm m}}
\newcommand{\Onu}{\Omega_{\nu}}
\newcommand{\onu}{\omega_{\nu}}
\newcommand{\lya}{Lyman-$\alpha$ }
\newcommand{\hMpc}{h/\mathrm{Mpc}}
\newcommand{\mnu}{m_{\nu}}
\newcommand{\Mnu}{M_{\nu}}
\newcommand{\eV}{\mathrm{eV}}
\def\vp{\varphi}
\def\bx{{\bf{x}}}
\def\bp{{\bf{p}}}
\def\bq{{\bf{q}}}
\def\bk{{\bf{k}}}

\input epsf

\preprint{LAPTH-1298/08, CERN-PH-TH/2008-248}
\title{Non-linear Power Spectrum including Massive Neutrinos: \\
the Time-RG Flow Approach} 

\author{
 Julien Lesgourgues,$^{1,2,3}$
Sabino Matarrese,$^{4,5}$
Massimo Pietroni,$^5$
Antonio Riotto$^{1,5}$
}
\vskip 0.5cm
\affiliation{
$^1$CERN, PH-TH Division, CH-1211, Geneve 23, Switzerland\\
$^2$Ecole Polytechnique F\'ederale de Lausanne, 
FSB/ITP/LPPC, BPS, Ch-10155, Lausanne, Switzerland\\
$^3$LAPTH, Universit\'e de Savoie, CNRS, B.P. 110, F-74941, 
Annecy-le-Vieux Cedex, France\\
$^4$Dipartimento di Fisica, Universit\'a di  Padova, Via Marzolo,
8 - I-35131 Padua - Italy\\
$^5$INFN, Sezione di Padova, Via Marzolo,
8 - I-35131 Padua - Italy
}
\date{\today}
\begin{abstract}
\noindent
Future large scale structure observations are expected to be sensitive
to small neutrino masses, of the order of 0.05 eV or more. However,
forecasts are based on the assumption that by the time at which these
datasets will be available, the non-linear spectrum in presence of
neutrino mass will be predicted with an accuracy at least equal to the
neutrino mass effect itself, i.e. about 3\%. Motivated by these
considerations, we present the computation of the non-linear power
spectrum of $\Lambda$CDM models in the presence of massive neutrinos
using the Renormalization Group (RG) time-flow approach, which amounts to a
resummation of perturbative corrections to the
matter power spectrum to all orders. 
We compare our results with those obtained with
other methods, i.e. linear theory, one-loop perturbation theory and
N-body simulations and show that the time-RG method improves
the one-loop method in fitting the N-body data, especially 
in determining the suppression of the  matter power spectrum when neutrino
are massive with respect to the linear power spectrum.  
\end{abstract}

\maketitle

\section{Introduction}
\noindent
The detection of atmospheric neutrino
oscillations proves that at least one neutrino mass eigenstate is as
heavy as approximately $(2.4 \pm 0.1 \times 10^{-3} \eV^2)^{1/2}$. In
the inverted hierarchy scenario, two eigenstates would have such a
large mass, while in the degenerate scenario, the three masses would
be of this order of magnitude or larger (for a recent analysis of the
data, see e.g.~\cite{Schwetz:2008er}). In all cases, the total mass
$\Mnu \equiv \sum \mnu$ should be greater than approximately 0.05 eV.
This means that neutrino masses are expected to impact significantly 
the cosmological evolution during matter and dark energy domination,
first by enhancing the background neutrino density (today, $\Onu =
(\Mnu / 93.14 \eV) h^{-2} \gsim 10^{-3}$, instead of $\Onu = 1.712
\times 10^{-5} h^{-2} $ in the massless limit) \cite{Mangano:2005cc},
and more importantly, by reducing the growth rate of matter
perturbations on comoving wavenumbers larger than $k_{\rm nr} \simeq 2
\times 10^{-3} \hMpc$. In the linear theory, this effect is maximal
for $k \gsim 1 \hMpc$ and suppresses the current matter power spectrum (PS)
by a factor $(1 - 8 \Onu / \Om)$, i.e. at least by 3\%
\cite{Hu:1997mj,Lesgourgues:2006nd}.

At the moment, the most robust cosmological bounds on the total
neutrino mass $\Mnu = \sum \mnu$ rely either on its background effect
(Komatsu et al. \cite{Komatsu:2008hk} obtain $\Mnu< 0.67 \eV$
[95\%C.L.] with WMAP-5, BAO and SNIa data only), or on limits on the
suppression of the matter PS in the linear regime (Tegmark et
al. \cite{Tegmark:2006az} obtain $\Mnu< 0.9 \eV$ [95\%C.L.] using the
combination of WMAP-3 with the PS of SDSS-LRG in the range $0.01 <
k<0.2 \hMpc$, after marginalization over an unknown nuisance parameter
describing non-linear effects in the range $0.1<k<0.2 \hMpc$). In
order to improve significantly these limits, it would be crucial to
model the effect of neutrino masses in the non-linear regime, in order
to include smaller scales in the analysis and increase the lever
arm. For galaxy redshift surveys, including wavenumbers above $k \gsim
0.1 \hMpc$ remains a delicate issue, because in addition to
non-linear corrections to the theoretical matter PS, one should be
able to predict the scale-dependence of the light-to-mass bias, which
starts to play a role on those scales. However, the prospects for
neutrino mass determination are particularly encouraging thanks to
other types of experiments.  For instance, future weak lensing surveys
(combined eventually with CMB lensing measurements) might reach a
sensitivity of $\Mnu \sim 0.04 \eV$ (2-$\sigma$) according to
\cite{Song:2003gg} (see also
\cite{Hannestad:2006as,Kitching:2008dp,Lesgourgues:2005yv,Perotto:2006rj}). 
These measurements -- which are not affected by any light-to-mass bias --
can probe the matter PS up to $z \sim 3$, which means, first, that they
are likely to detect the non-trivial redshift dependence of the
small-scale matter PS in the presence of massive neutrinos; and second,
that they can measure cosmological fluctuations deeper in the linear
regime than observations limited to small redshifts. Nevertheless, the
forecasts of \cite{Song:2003gg,Hannestad:2006as,Kitching:2008dp} do
include non-linear scales, corresponding to multipoles $\ell < 1000$
in the case of \cite{Song:2003gg} (for comparision, non-linear effects
are important at $\ell > 40$ for sources with $z_s \sim 0.2$, and
$\ell>200$ for $z_s \sim 3$). Hence, they are based on the assumption
that by the time at which these datasets will be available (of the
order of one decade), the non-linear spectrum in the presence of neutrino
mass will be predicted with an accuracy at least equal to the neutrino
mass effect itself, i.e. about 3\%.
The same comment applies to other promising 
forecasts related to future cluster surveys \cite{Wang:2005vr}, 21cm
surveys \cite{Wyithe:2008mv,Pritchard:2008wy}, 
high-redshift galaxy surveys \cite{Hannestad:2007cp}, 
CMB-weak lensing
cross-correlation \cite{Ichikawa:2005hi,Lesgourgues:2007ix} or \lya
forest data \cite{Gratton:2007tb}. In the latter case, one should
model the effect of neutrino mass not only on the total matter power
spectrum, but also on the thermodynamical evolution of the
intergalactic medium: hence, massive neutrinos should be implemented
in hydrodynamical simulations of structure formation before deriving
robust bounds on neutrino masses. All previous neutrino mass bounds
from \lya forest data neglected this issue.

Current weak lensing data probe the matter PS on much
smaller scales than the previously discussed experiments (typically of
the order of $k \sim (10-40)\, \hMpc$). In this case, neutrino mass bounds
have been obtained using the {\sc halofit} approach in order to
compute non-linear corrections to the CDM component of the total power
spectrum \cite{Tereno:2008mm,Ichiki:2008ye}.  The {\sc halofit} algorithm was
derived and calibrated for the $\Lambda$CDM model. Its use in presence
of massive neutrinos might be correct, but
this would need further justification (see \cite{Abazajian:2004zh}).

Finally, the computation of the non-linear PS in presence
of massive neutrinos would be important for CMB physics. A crucial
issue in CMB observations is to deal with foreground contamination. In
particular, the thermal SZ effect is difficult to remove, and a large
fraction of it will contribute to the final measured temperature
spectrum. This SZ effect depends on the non-linear PS;
hence, it can be computed and fitted at the same time as primary
anisotropies.  However, any mistake in the computation of this
contribution would result in a significant bias on the cosmological
parameters (for instance, on the scalar spectral tilt)
\cite{Dore}. Even with the smallest allowed neutrino mass, the SZ
spectrum should be sensitive to neutrino free-streaming on small
scales. So, if the CMB spectrum of a precise experiment like Planck
was analysed using SZ spectrum computations in which neutrino mass
effects are neglected, the bias induced on cosmological parameters
might be significant. Note that instead of fitting this component, one
can marginalize over nuisance parameters accounting for its amplitude
and slope, or just restrict the multipole range used in the analysis,
but this is at the expense of reducing the parameter sensitivity
\cite{Dore}. Hence, investigating the effect of neutrino masses on
non-linear cosmological perturbations is also important in order to
push the precision of CMB observations on all comsmological
parameters.

For all these reasons, a crucial step for precision cosmology is to
incorporate free-streaming massive neutrinos in numerical or
analytical estimates of the non-linear matter PS on small
scales. On the side of numerical simulations, the first steps were performed in
\cite{Klypin,Primack,Brandbyge:2008rv,Brandbyge:2008js}. 
To our knowledge, other attempts to simulate mixed cold
plus hot models were treating neutrinos as extra cold particles with a
different initial PS. This approach neglects the effect of
neutrino thermal velocities, and fails to reproduce even the linear
evolution. Ref.~\cite{Brandbyge:2008rv} shows explicitely that
neglecting thermal velocities results in an error which is of the same
size as the very effect of a small neutrino mass: this makes current
bounds on $\mnu$ from \lya data suspicious. However, implementing
free-streaming neutrino ``particles'' in simulation is very demanding,
because of the large number of particles required to sample
the full neutrino phase space, and of the tiny time steps needed to
follow the particle trajectories with sufficient accuracy, because of
their large velocities. Some
significant progress can be expected in this field in the next years.

In parallel, it is important to derive analytic approximations of the
non-linear evolution in order to cross-check the results from
simulations, to reach eventually smaller scales and larger redhifts,
and hopefully to obtain computationally faster ways of predicting
non-linear spectra. Various analytical schemes have been discussed in
the past years for $\Lambda$CDM models, inspired by quantum field
theory; these include one-loop calculations \cite{PTreview,JK06, JK08}, 
resummation techniques \cite{CS1, CS2, Taruya2007},  renormalization group aproaches 
\cite{McD06, MP07a, MP07b,Izumi07,Matsubara07,ValBisp,Matsubara:2008wx}, 
or the Time Renormalization Group (TRG) flow calculation of \cite{P08a}.  

The first attempt to incoporate massive
neutrinos in these calculations was performed by Saito et
al. \cite{Saito:2008bp}, who provide a first-order approximation of
the one-loop computation. The full one-loop results were obtained by Wong
\cite{Wong:2008ws}. The  goal of this paper is to generalize the method of
\cite{P08a} in the presence of neutrino masses, and to compare with previous
results from \cite{Brandbyge:2008rv,Saito:2008bp,Wong:2008ws}. 
In this  new approach the matter power spectrum, the bispectrum ({\it i.e.}
the connected three point-correlation function in Fourier space) 
and higher order correlations, are obtained -- at any redshift and 
for any momentum scale -- 
by integrating a system of differential equations.
Truncating at the level of the trispectrum  ({\it i.e.}
the connected four point-correlation function in Fourier space), 
the solution of the equations corresponds to the summation of an 
infinite class of perturbative corrections. 
In the diagrammatic language, this is equivalent to resumming at all 
orders in perturbation theory the corrections
to the matter power spectrum.  Furthermore, comparison with N-body
simulations for zero neutrino masses shows that the method is valid
up to momentum wavenumbers of the order of 0.3 $h$/Mpc at $z=1$, and able to describe
the non-linearities due to mode-to-mode coupling in the
baryon acoustic oscillations region. 
Compared to other resummation frameworks, this scheme is particularly 
suited to cosmologies other than $\Lambda$CDM, 
such as those based on modifications of gravity and those containing 
massive neutrinos. 

The paper is organized as follows: in Section 1 we summarize our method 
based on Ref. \cite{P08a} generalized to include
massive neutrinos. In Section 3 we present our numerical findings and 
compare them with those obtained with other methods, 
namely linear theory, one-loop perturbation theory 
\cite{Saito:2008bp,Wong:2008ws} and the N-body simulations of
Refs. \cite{Brandbyge:2008js}. 

\section{Method}
\noindent
Our goal is to set up a method to compute non-linear corrections to the 
matter PS in a mixed dark matter (MDM) model. The matter density fluctuation 
is given by 
\beq \delta_m\equiv(\delta \rho_c +\delta 
\rho_b+\delta \rho_\nu)/\bar{\rho}_m = (1-f_\nu)\delta_{cb}+f_\nu \delta_\nu\,,
\label{dm}
\eeq
where the subscripts `$c$', `$b$', `$\nu$', and `$cb$', 
stand for cold dark matter (CDM), baryons, neutrinos, and CDM plus baryons, 
respectively, and $f_\nu=\Omega_\nu/\Omega_m$. The total matter PS is 
then given by
\beqra
&& P_m(k; z) =(1-f_\nu)^2\, P_{cb}(k; z)+ 2\, (1-f_\nu) f_{\nu} 
\,P_{cb,\nu} (k; z)+  f_{\nu}^2 \, P_{\nu} (k; z).\label{PSTOT}
\eeqra 
In ref.~\cite{Saito:2008bp} the expression above was approximated 
by using the linear order result for the 
$cb-\nu$ and the $\nu-\nu$ PS's (the second and third terms in 
Eq.~(\ref{PSTOT}), respectively), while including non-linear corrections 
to the $cb-cb$ one in the so-called one-loop approximation, which 
includes density and velocity perturbations up to the third order 
\cite{PTreview}.
In the present paper we will go beyond the one-loop approximation 
for $P_{cb}$, using the TRG approach of Ref.~\cite{P08a}, 
while keeping the linear order results for the other two contributions 
to Eq.~(\ref{PSTOT}). 

The TRG is a method to sum perturbative corrections to all orders. 
The starting point are the continuity and Euler equations satisfied 
by the density contrast and peculiar velocity of a non-relativistic fluid 
({\it i.e.} CDM or baryons in the present case),
\beqra
&&\frac{\partial\,\delta_{cb}}{\partial\,\tau}+
{\bf \nabla}\cdot\left[(1+\delta_{cb}) {\bf v} \right]=0\,,\nonumber\\
&& \frac{\partial\,{\bf v}}{\partial\,\tau}+{\cal H}\,{\bf v}\, + ( {\bf v} 
\cdot {\bf \nabla})  {\bf v}= - {\bf \nabla} \phi\,,
\label{Euler}
\eeqra
where $\tau$ is the conformal time and we have assumed 
$\delta_b=\delta_c=\delta_{cb}$. 
The gravitational potential $\phi$ is determined by the total mass 
fluctuation, via the Poisson equation
\beq
\nabla^2 \phi = \frac{3}{2}\,\,{\cal H}^2  \,\Omega_m \, \delta_m\,,
\label{Poisson}
\eeq
where ${\cal H}= d \log a/d\tau$. Dropping the $\delta_{cb}{\bf v}$ term in 
the continuity equation and the $ ({\bf v}  \cdot {\bf \nabla}) {\bf v}$ 
one in the Euler equation, the solutions to the system above reproduce 
linear perturbation theory on the subhorizon scales we are interested in. 

In order to close the system (\ref{Euler}), we go to Fourier space, use 
Eq.~(\ref{dm}), and approximate the RHS of the Poisson equation as follows
\beq
\frac{3}{2}\,\,{\cal H}^2  \,\Omega_m(\tau) \, 
\delta_m(\bk,\,\tau)\,\simeq \frac{3}{2}\,\,{\cal H}^2  
\,\Omega_{cb}^{eff}(\bk,\,\tau) \,\delta_{cb}(\bk,\,\tau)\,,
\eeq
where 
\beq
\label{aa}
\Omega_{cb}^{eff}(\bk,\,\tau) \equiv \Omega_m(\tau) (1-f_{\nu})
\left(1+\frac{f_\nu\delta_\nu^L(\bk,\,\tau)}{(1-f_{\nu})
\delta_{cb}^L(\bk,\,\tau)} \right)\,.  \eeq $\delta_{\nu,cb}^L$
indicate the density perturbations evolved according to linear
theory. Notice that, due to the different space
dependence of $\delta_\nu^L$ and $\delta_\nu^{cb}$ in the massive
neutrino case, $\Omega_{cb}^{eff}$ -- unlike $\Omega_m$ -- is
space-dependent. Another approximation is to recursively use the full
TRG-evolved $\delta_{cb}$ fluctuation in Eq.~(\ref{aa}). However, we
numerically checked that performing one iteration (i.e., using the
output non-linear $\delta_{cb}$ as an input in Eq.~(\ref{aa}) instead
of $\delta_{cb}^L$) does not affect the results by more than 0.1\%.
This way of including massive neutrinos is conceptually similar to
the grid method used in N-body simulations of Ref.~\cite{Brandbyge:2008js}. where,
instead of simulating massive neutrinos as particles with individual 
velocities, they are embedded as a local neutrino density on a grid which is 
evolved in time using linear theory. 

Next, we  write Eqs.~(\ref{Euler}, \ref{Poisson}) in a compact form 
\cite{CS1,P08a}. First, we introduce the doublet $\vp_a$ ($a=1,2$), given by
\beq\left(\begin{array}{c}
\varphi_1 ( {\bf k}, \eta)\\
\varphi_2 ( {\bf k}, \eta)  
\end{array}\right)
\equiv 
e^{-\eta} \left( \begin{array}{c}
\delta_{cb}  ( {\bf k}, \eta) \\
-\theta  ( {\bf k}, \eta)/{\cal H}
\end{array}
\right)\,,
\label{doppietto}
\eeq
where $\theta = i\, {\bf k}\cdot {\bf v}$, and the time variable 
has been replaced by the logarithm of the scale factor,
\beq
\eta= \log\frac{a}{a_{in}}\,,
\label{etadef}
\eeq
$a_{in}$ being the scale factor at a conveniently remote epoch, 
such that all the relevant scales are well inside the linear regime. 

Then, we get 
\beqra
&& \partial_\eta\,\varphi_a({\bf k}, \eta)= -\Theta_{ab}({\bf k},\,\eta )
\varphi_b({\bf k}, \eta) + e^\eta 
\gamma_{abc}({\bf k},\,-{\bf p},\,-{\bf q})  
\varphi_b({\bf p}, \eta )\,\varphi_c({\bf q}, \eta ),
\label{compact}
\eeqra
where 
\beq{\bf \Theta} ({\bf k},\,\eta ) = \left(\begin{array}{cc}
\ds 1 & \ds -1\\
\ds -\frac{3}{2} \Omega_{cb}^{eff}(\bk,\,\eta) & \ds 2 +
\frac{d \log{\cal H}}{ d \eta} \end{array}
\right)\,,
\label{bigomega}
\eeq
and the only non-vanishing elements of the {\it vertex} function 
$\gamma_{abc}({\bf k},{\bf p},{\bf q}) $  are
\beqra
&&\gamma_{112}({\bf k},\,{\bf p},\,{\bf q}) = 
\frac{1}{2} \,\delta_D ({\bf k}+{\bf p}+{\bf q})\, 
\frac{(\bp + \bq) \cdot \bq}{q^2}\,\,,\nonumber\\
&&\gamma_{222}({\bf k},\,{\bf p},\,{\bf q}) = 
\delta_D ({\bf k}+{\bf p}+{\bf q})\, \frac{(\bp + \bq)^2 \,
\bp \cdot \bq}{2 \,p ^2 q^2}\,,
\label{vertice}
\eeqra
and 
$\gamma_{121}({\bf k},\,{\bf p},\,{\bf q})  = 
\gamma_{112}({\bf k},\,{\bf q},\,{\bf p}) $.

In Eq.~(\ref{compact}), repeated indices are summed over, and integration 
over momenta $\bq$ and $\bp$ is understood.
Notice that all the information on the neutrino mass, both at the background 
and at the linear perturbation level, is contained in $\Theta_{21}$, 
the other entries of ${\bf \Theta} $ and the vertices being universal.

At this point we should stress a crucial difference between the present approach and the ones of refs.~\cite{CS1,MP07a,MP07b,CS2}. Namely, in those papers, the equations are derived in the Einstein-deSitter (EdS) case ($\Omega_m=1$) and then extended to different cosmologies, such as $\Lambda$CDM, by reinterpreting $\eta$ as the logarithm of the linear growth factor, while keeping the equations unchanged. In particular, the $\bf{\Theta}$ - matrix is approximated by that of the EdS model. It was discussed in \cite{CS1} and numerically checked in \cite{P08a} that in cosmologies with a constant equation of state for the dark energy, this procedure gives at most a $O(1\%)$ error at $z=0$ for $k\gsim 0.3 h$/Mpc, rapidly decreasing at higher redshift and larger scales. The physical reason for this accuracy lies in the fact that this procedure takes fully into account the growing mode of the non-EdS cosmology, while it mistreats the decreasing one. The latter start to play a role only at high $k$'s or for low redshifts, where non-linearities become important.

However, when massive neutrinos contribute to the dark matter, the linear growth factor is $k$-dependent, and the redefinition of $\eta$ is ill-defined. This singles out the present approach, the TRG, as particularly suited to this case. Indeed, in this approach, we keep the same definition, Eq.~(\ref{etadef}) for any cosmological model, and take fully into account the $k$-dependence of the linear growth function via the $\Omega^{eff}_{cb}({\bf k},\,\eta)$ term in Eq.~(\ref{bigomega}).

The evolution in $\eta$ of the correlation functions of the $\vp_a$ fields 
can be derived by iterating the application of Eq.~(\ref{compact}). 
The result is an infinite tower of coupled integro-differential equations. 
Following Ref.~\cite{P08a}, we will truncate the system by neglecting the 
trispectrum. The evolution equations for the PS, $\langle 
\varphi_a({\bf k},\,\eta) \varphi_b({\bf q},\,\eta)\rangle \equiv 
\delta_D({\bf k + q}) P_{ab}({\bf k}\,,\eta)$, and for the bispectrum, 
$\langle\varphi_a({\bf k},\,\eta) 
\varphi_b({\bf q},\,\eta)\varphi_c({\bf p},\,\eta)\rangle 
\equiv \delta_D({\bf k + q+p})
B_{abc}({\bf k},\,{\bf q},\,{\bf p};\,\eta)$ are then given by
\beqra
\ds  \partial_\eta\,P_{ab}({\bf k}\,,\eta) &=& - 
\Theta_{ac} ({\bf k}\,,\eta)P_{cb}({\bf k}\,,\eta) - 
\Theta_{bc} ({\bf k}\,,\eta)P_{ac}({\bf k}\,,\eta) \nonumber\\
&&+\,e^\eta \int d^3 q\, \left[ 
\gamma_{acd}({\bf k},\,{\bf -q},\,{\bf q-k})\,B_{bcd}({\bf k},\,{\bf -q},\, 
{\bf q-k};\,\eta)\right.\nonumber\\
&&\left. + \,B_{acd}({\bf k},\,{\bf -q},\,{\bf q-k};\,\eta)\,
\gamma_{bcd}({\bf k},\,{\bf -q},\,{\bf q-k})\right]\,,\nonumber\\
&&\nonumber\\
\ds \partial_\eta\,B_{abc}({\bf k},\,{\bf -q},\,{\bf q-k};\,\eta) &=&  
- \Theta_{ad} ({\bf k}\,,\eta)B_{dbc}({\bf k},\,{\bf -q},\,{\bf q-k};\,\eta) 
- \Theta_{bd} ({\bf -q}\,,\eta)B_{adc}({\bf k},\,{\bf -q},\,{\bf q-k};\,\eta)
\nonumber\\
&& \;- \Theta_{cd} ({\bf q-k}\,,\eta) 
B_{abd}({\bf k},\,{\bf -q},\,{\bf q-k};\,\eta)\nonumber\\
&& \;+ 2 e^\eta \left[ \gamma_{ade}({\bf k},\,{\bf -q},\,{\bf q-k}) 
P_{db}({\bf q}\,,\eta)P_{ec}({\bf k-q}\,,\eta)\right.\nonumber\\
&& \;+\gamma_{bde}({\bf -q},\,{\bf q-k},\,{\bf k}) 
P_{dc}({\bf k-q}\,,\eta)P_{ea}({\bf k}\,,\eta)\nonumber\\
&&\;+\left. \gamma_{cde}({\bf q-k},\,{\bf k},\,{\bf -q}) 
P_{da}({\bf k}\,,\eta)P_{eb}({\bf q}\,,\eta)\right]\,.
\label{syst}
\eeqra 
The system is solved by giving initial conditions for the PS
and the bispectrum at $\eta=0$, and integrating forward in $\eta$. In
the $a_{in}\to 0$ limit, the initial conditions can be obtained from
linear perturbation theory, \beqra &&P_{11}(\bk,\,0)=
P_{11}^L(\bk,\,0)\,,\nonumber \\ &&P_{12}(\bk,\,0)=f_{L} (\bk,0)
P_{11}^L(\bk,\,0)\,,\nonumber\\ &&P_{22}(\bk,\,0)=f_{L}^2(\bk, 0)
P_{11}^L(\bk,\,0)\,,\nonumber\\ &&B_{abc}({\bf k},\,{\bf -q},\,{\bf
q-k};\,0) =0\,, \eeqra where the multiplication of the initial
$P_{12}$ and $P_{22}$ by appropriate powers of the linear growth
function, $f_L(\bk,\,\eta) \equiv d\log \delta_{cb}^L(\bk,\,\eta)/d
\eta$, selects the linear growing mode.  In our computations, we will
use $a_{in}=1/41$, and take the density-density power spectrum
$P_{11}^L$ from the Boltzman code {\sc camb} \cite{Lewis:1999bs}. 
More precisely,
we extract $P_{cb}$, $P_{cb, \nu}$, $P_{\nu}$ and $\Omega_{cb}^{eff}$ at
any value of $(k,z)$ from the quantities $\delta_{cb}$ and
$\delta_\nu$ evolved by {\sc camb} in the synchronous gauge.

Different approximations of the system above give the linear and the one-loop approximations, respectively. 
Indeed, the linear result for the PS is obtained simply by setting 
$B_{abc}=0$, and integrating the first of Eqs.~(\ref{syst}). 
On the other hand, the one-loop result is obtained by inserting the 
linear PS in the RHS of the second of Eqs.~(\ref{syst}) and using 
the resulting $B_{abc}$ in the first one. 

Integrating the system above as it is, goes beyond the one-loop approximation, 
and corresponds to summing infinite perturbative contributions. 
As discussed in detail in Ref.~ \cite{P08a}, at each step in $\eta$ the 
PS and the bispectrum appearing at the RHS's include non-linear corrections 
maturated from $\eta=0$ to $\eta$. 

In the next section we will present the results of such integrations 
and compare them with the linear, the one-loop approximations and the 
N-body results.

\section{Results and discussion}
\begin{figure}
\centering
\includegraphics[width = 0.5\textwidth,keepaspectratio=true]{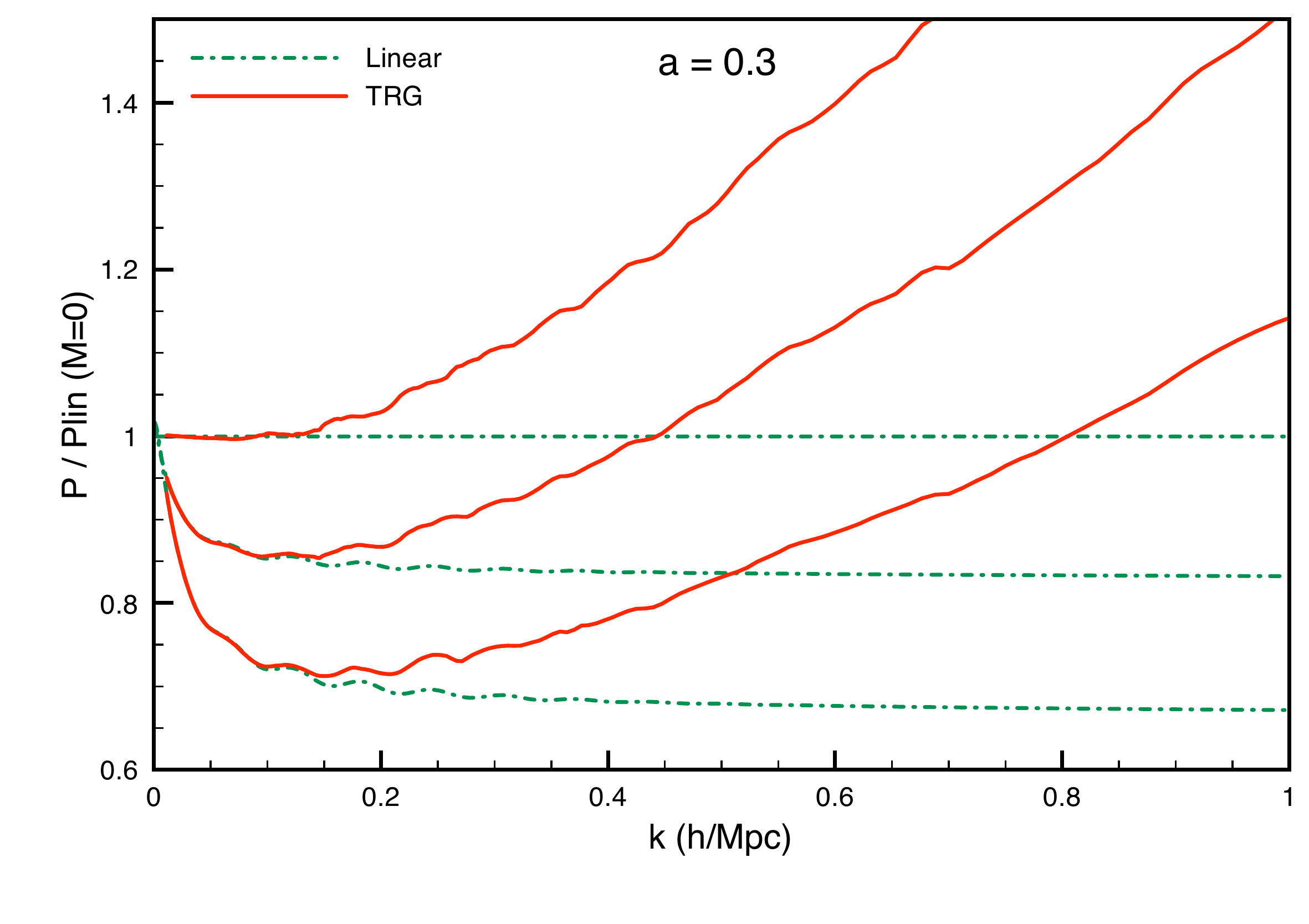}\hfill
\caption{Matter power spectra at $a=0.3$ ($z=2.33$) normalized to the linear one for $M_\nu=0$. The red, solid lines are the result of the TRG method described in the text, while the green, dash-dotted ones are the linear approximation. The curves from top to bottom correspond to $M_\nu =0,\,0.3,$ and $0.6$ eV, respectively.}
\label{tutto}
\end{figure}

\noindent
The impact of non-linear corrections on the linear PS is opposite with respect to that of a massive neutrino component. Indeed, while the latter damps the PS at scales larger than $k_\mathrm{nr} \simeq 2 \times 10^{-3} h/\mathrm{Mpc}$, non-linearities enhance it for $k \gsim k_\mathrm{NL}$, where $k_\mathrm{NL} \simeq 0.05  h/\mathrm{Mpc}$ at $z=0$ and scales roughly as $(1+z)^{2/3}$. The combination of both effects is shown in Fig.~\ref{tutto}, where we plot the total matter PS normalized to the linear one for $M_\nu=0$, at $a=0.3$, {\it i.e.} $z=2.33$ ($a$ is the cosmological scale factor, normalized to $a=1$ today). We have assumed a flat cosmological model with Hubble parameter $h=0.70$, and density parameters $\Omega_\Lambda=0.7$ and $\Omega_m=0.3$. $\Omega_m$ is given by the sum of $\Omega_b=0.05$ and $\Omega_c+\Omega_\nu$. The primordial power spectrum was assumed of the scale-invariant Harrison-Zel'dovich form ($n=1$), normalized as to give $\sigma_8=0.878$ for  $\Omega_\nu=0$.
 
The green, dash-dotted lines are the linear approximation result, while the red ones were obtained with the TRG method. The curves from top to bottom correspond to $M_\nu =0,\,0.3,$ and $0.6$ eV, respectively. As we see, even at an intermediate redshift the effect of non-linearities overcomes that of massive neutrinos, turning a $k$-independent suppression in a $k$-dependent enhancement, which should be understood in order to derive reliable bounds on neutrino masses.
 
\begin{figure}
\includegraphics[width = 0.33\textwidth,keepaspectratio=true]{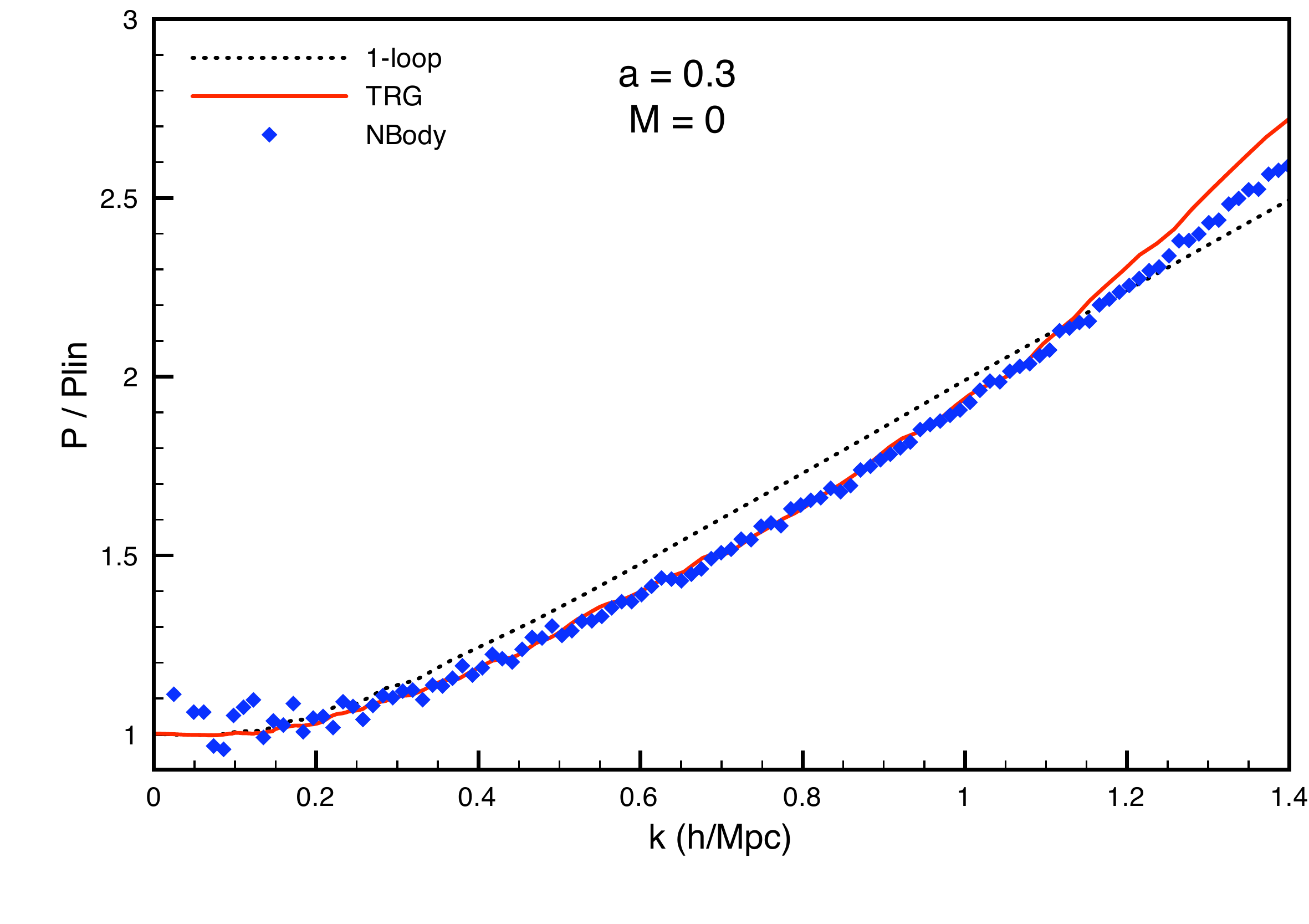}\hfill
\includegraphics[width = 0.33\textwidth,keepaspectratio=true]{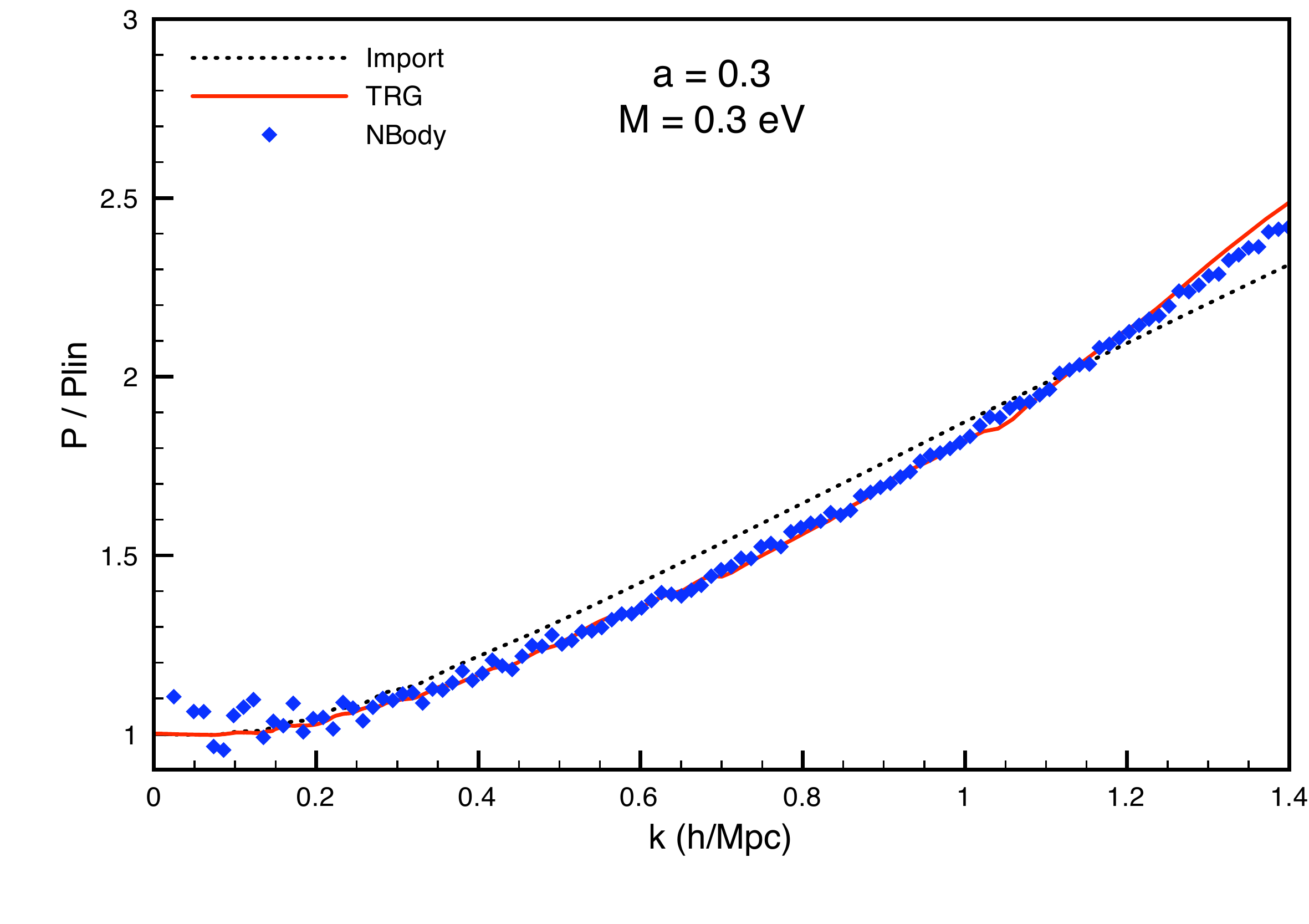}\hfill
\includegraphics[width = 0.33\textwidth,keepaspectratio=true]{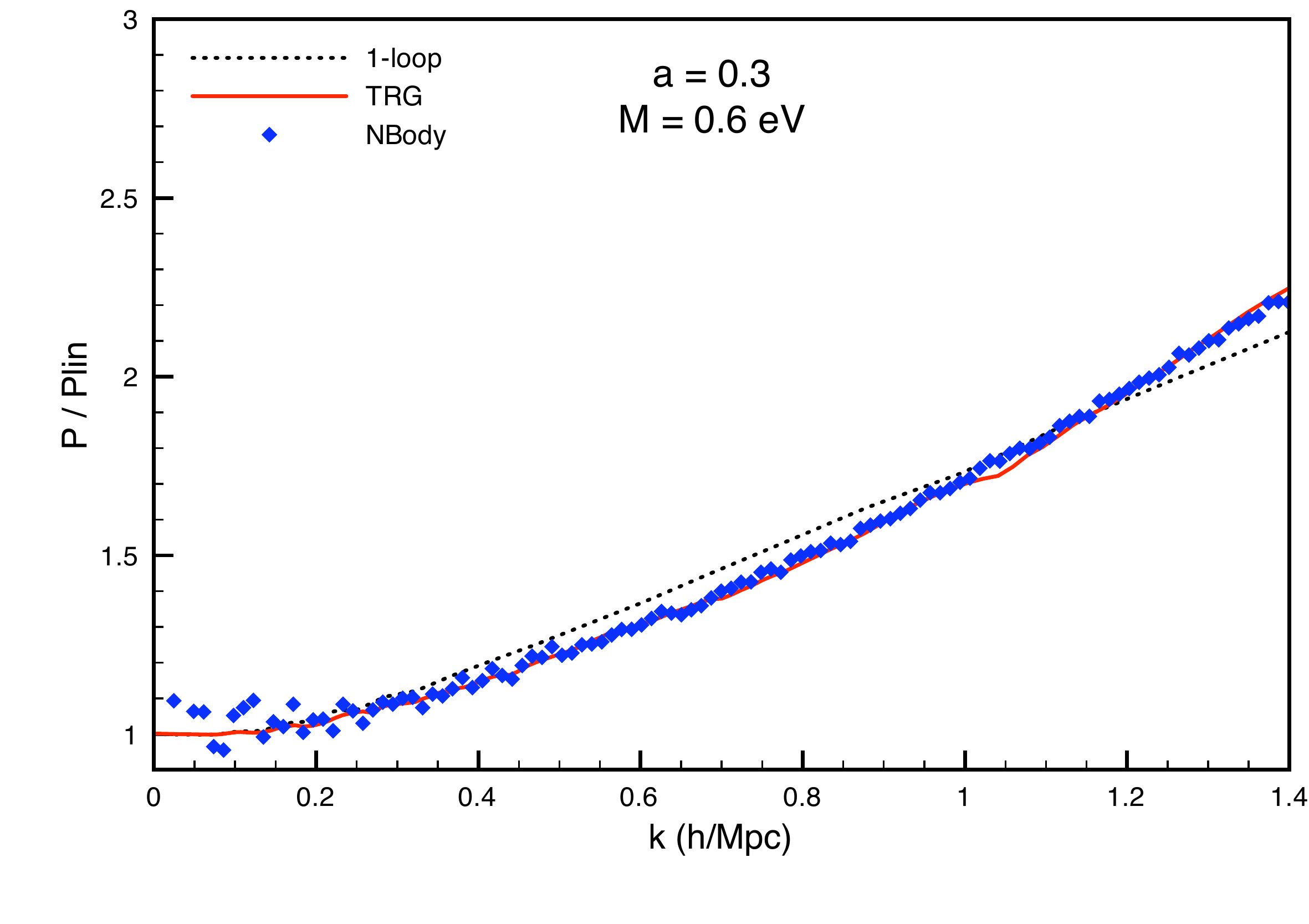}\\
\includegraphics[width = 0.33\textwidth,keepaspectratio=true]{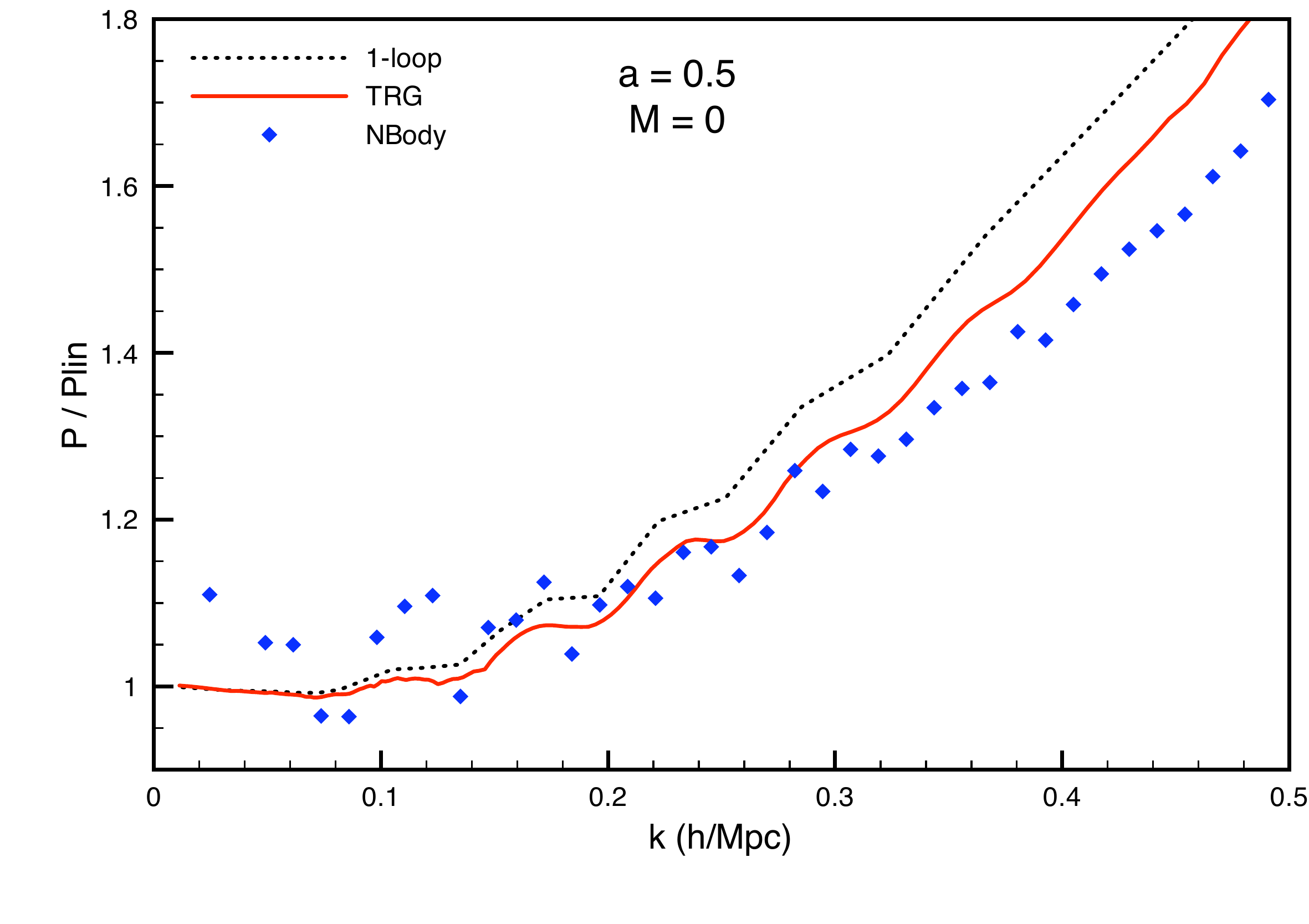}\hfill
\includegraphics[width = 0.33\textwidth,keepaspectratio=true]{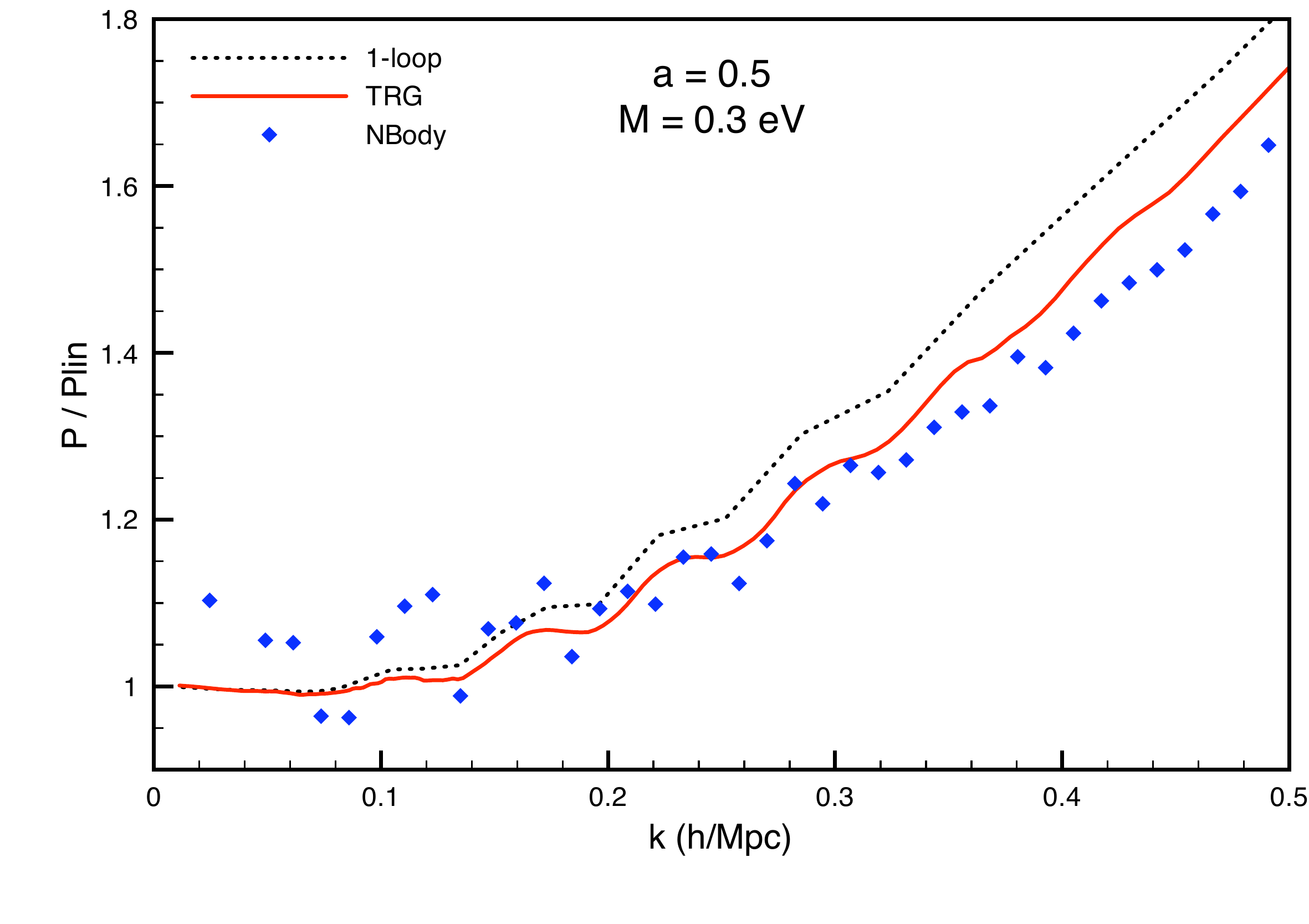}\hfill
\includegraphics[width = 0.33\textwidth,keepaspectratio=true]{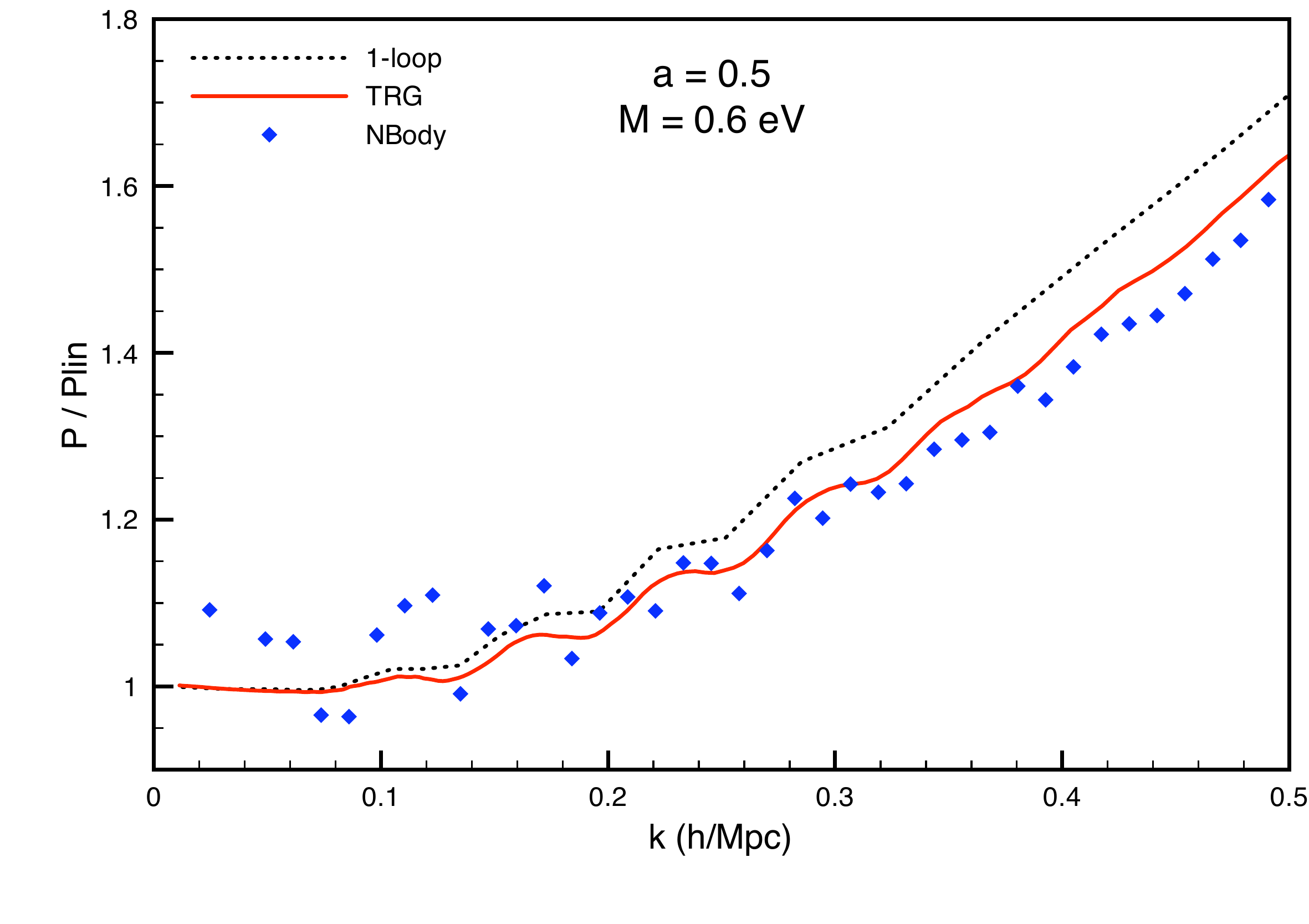}\\
\caption{Power spectra divided by the linear one for the same value of 
$M_\nu$ as computed in one-loop perturbation theory (black, dotted), and by the TRG method 
(red, solid). Blue diamonds are the results of  N-body simulation of Ref.~\cite{Brandbyge:2008js}, when
neutrinos are treated like particles.}
\label{RatioAbs}
\end{figure}

In the following figures  we disentangle the non-linear from the massive neutrinos effects. 
Non-linear effects are shown in Fig.~\ref{RatioAbs}, where the total matter power spectra is divided by the linear one computed for the same value of the total neutrino mass. We compare results for  $M_\nu=0, 0.3$ and 0.6 eV obtained in one-loop perturbation theory 
(black dotted), and by the TRG method 
(red solid).  Blue diamonds are the results of  N-body simulations of Ref. \cite{Brandbyge:2008js}, when
neutrinos are treated like particles. The results are
given for $a=0.3$ corresponding to redshift $z=2.33$ and $a=0.5$ 
corresponding to  $z=1$. 

At high redshift, the TRG method works surprisingly well in the
whole range of $k$ displayed in the figure, and is able to reproduce
the N-body simulations better than the one-loop result, especially for
large neutrino masses.  For smaller redshifts, this tendency remains,
but as expected, the agreement between the N-body simulation and the
TRG method is valid only up to $k=0.3 h/$Mpc, while in the same range the 
one-loop method overestimates the non-linear to linear power spectra ratio.
At both redshifts, the agreement with N-body simulations increases with the neutrino mass. This can be understood by recalling that we keep $\Omega_c+\Omega_\nu$ fixed, and non-linear effects decrease when the CDM component is lowered. 

\begin{figure}
\centering
\includegraphics[width = 0.5\textwidth,keepaspectratio=true]{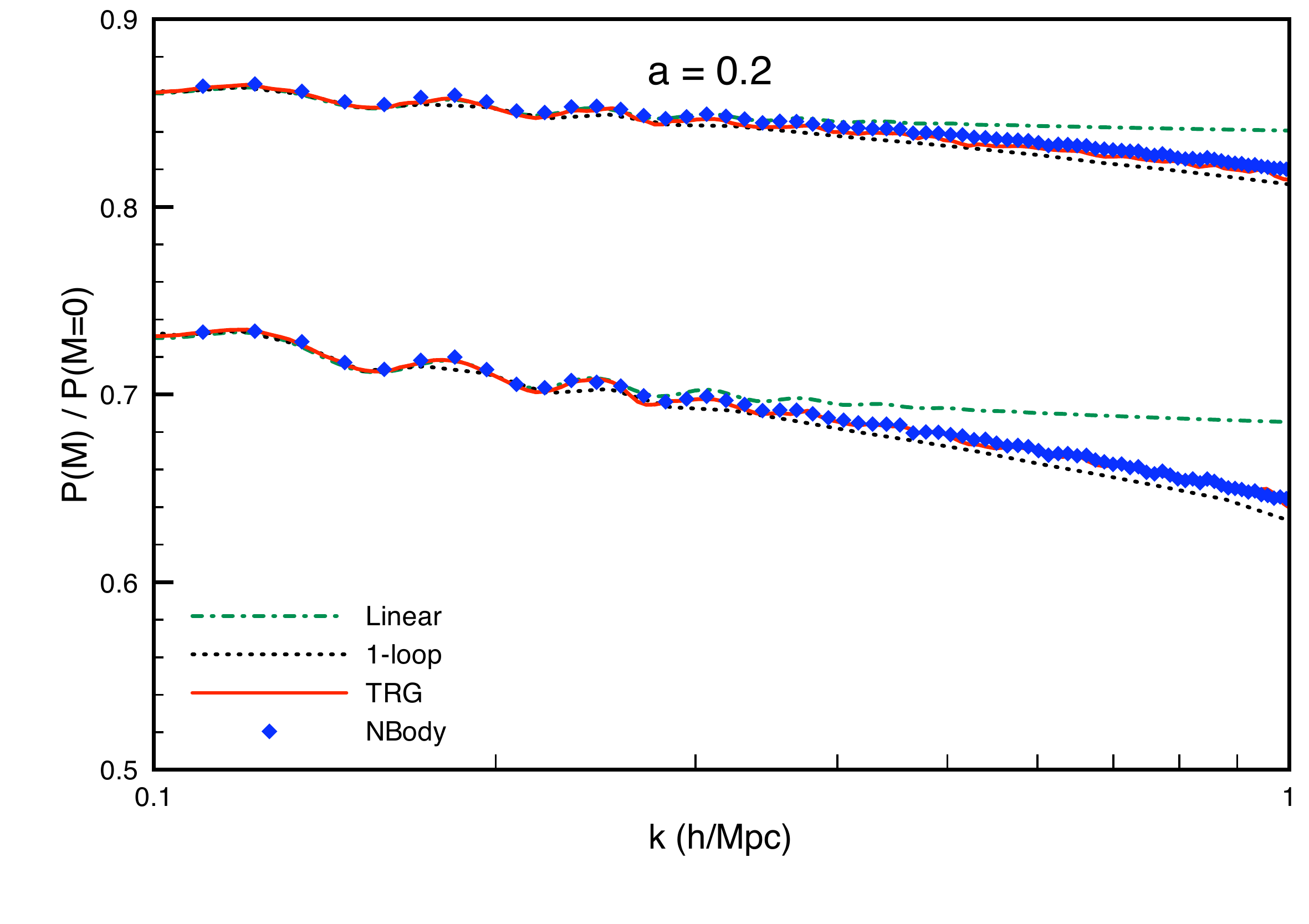}\hfill
\includegraphics[width =0.5\textwidth,,keepaspectratio=true]{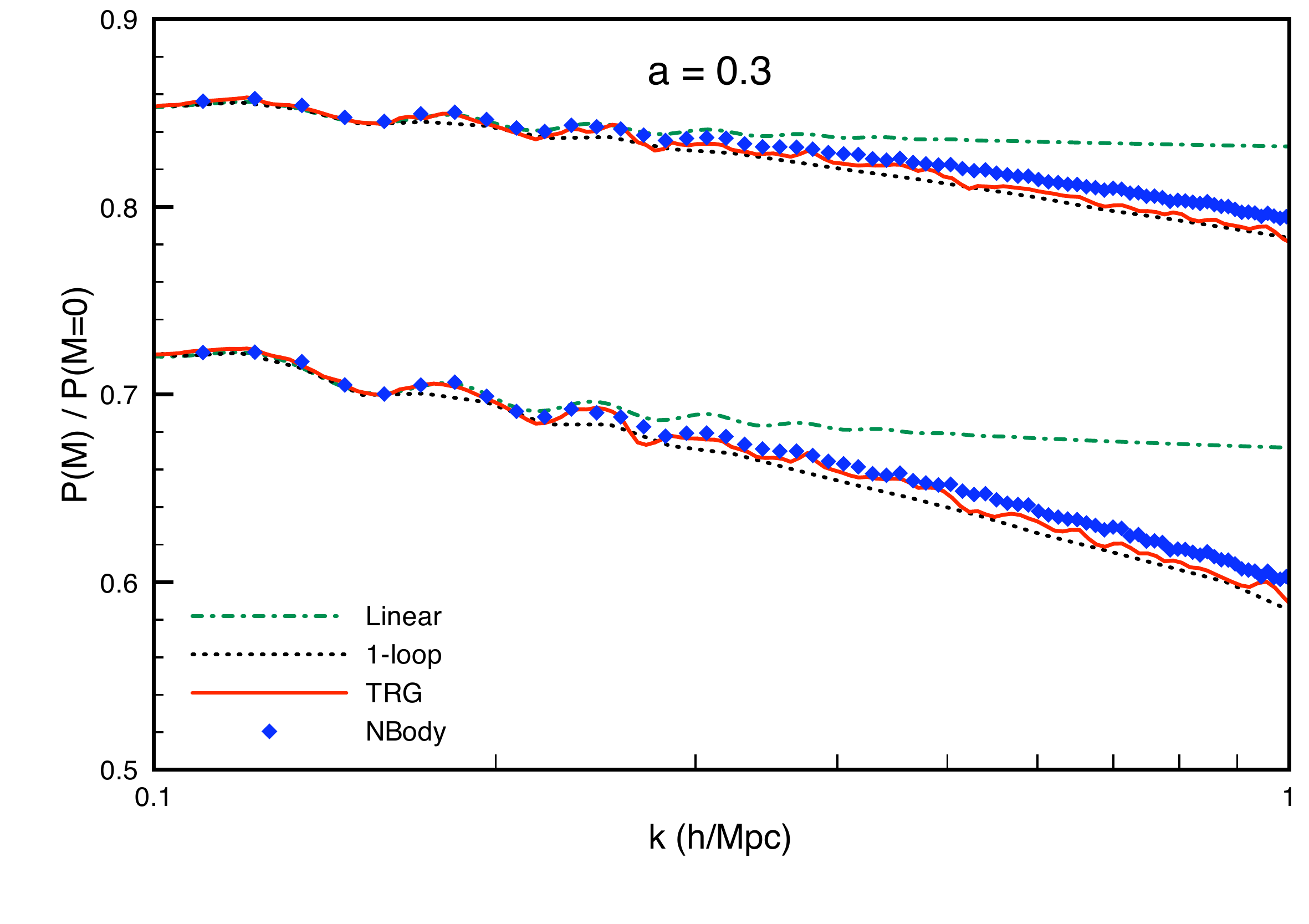}\\
\includegraphics[width = 0.5\textwidth,keepaspectratio=true]{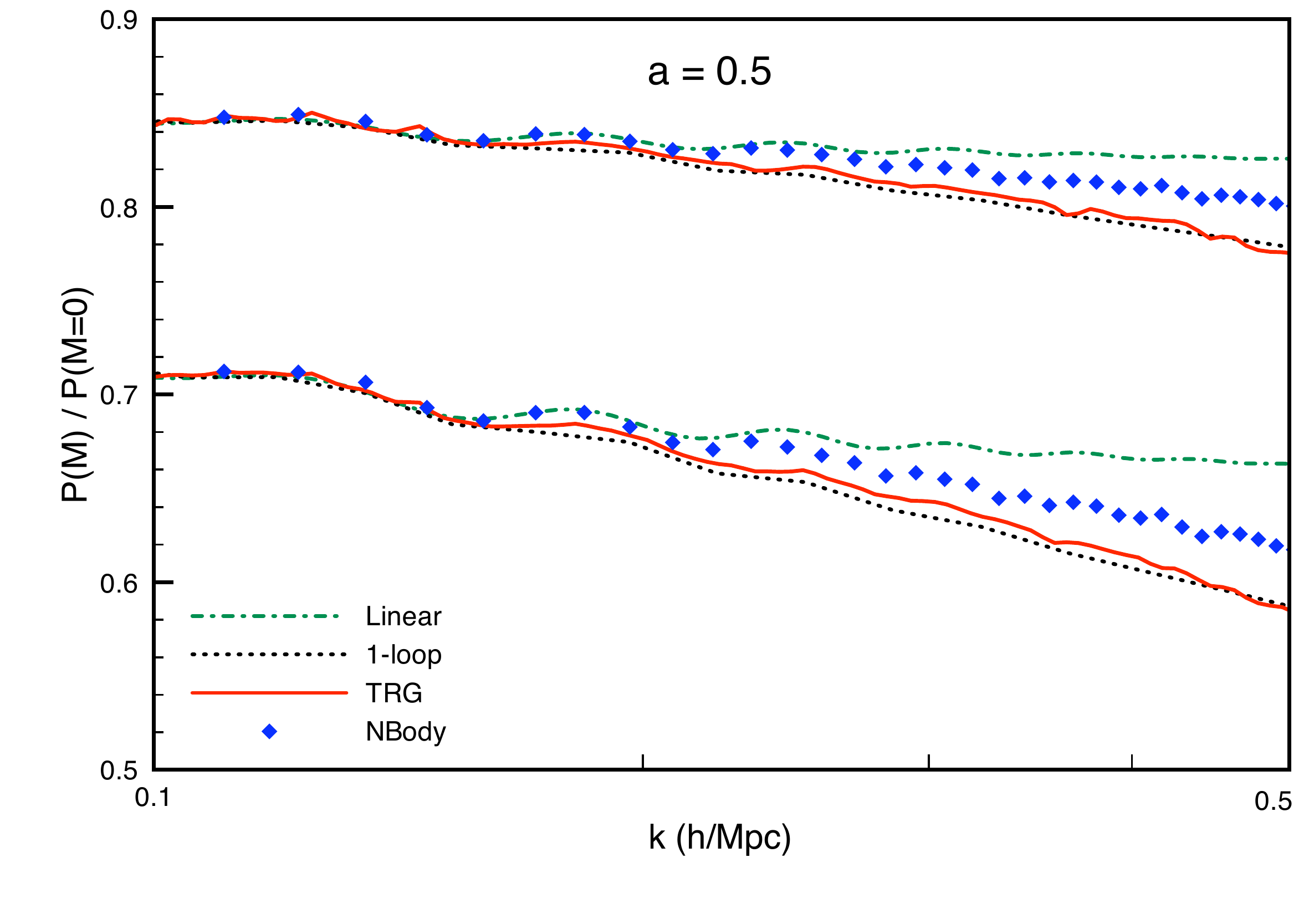}\hfill
\includegraphics[width = 0.5\textwidth,keepaspectratio=true]{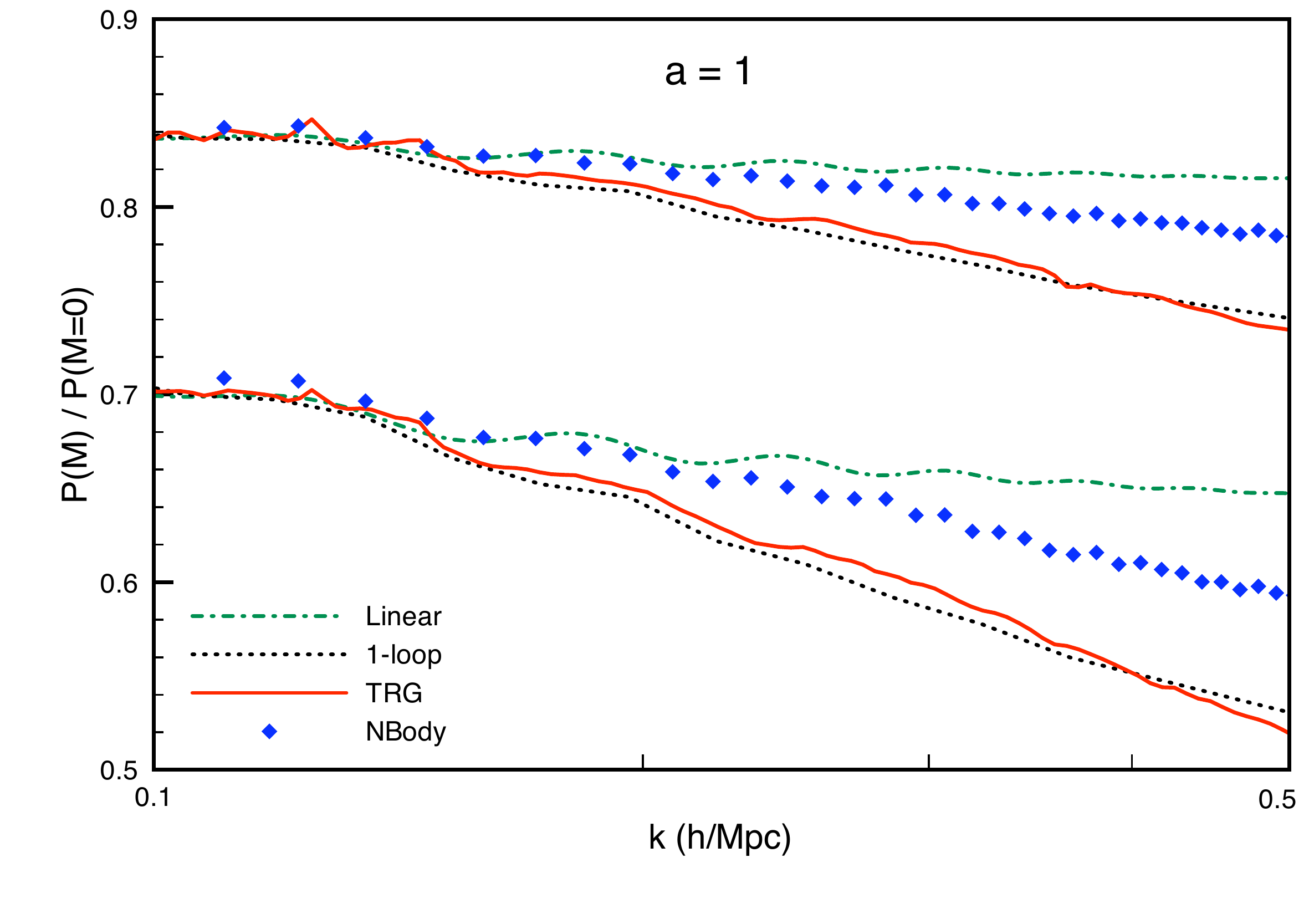}\\
\caption{Suppression on the total PS induced by non-zero neutrino masses at different epochs,
as computed in linear approximation (green, dash-dotted line), 
one-loop perturbation theory (black, dotted), and by the TRG method 
(red, solid). Blue diamonds are the results of  N-body simulation of Ref. \cite{Brandbyge:2008js}.The upper (lower) group of lines corresponds to  $M_\nu = 0.3$ ($0.6$) eV, respectively.}
\label{Ratios}
\end{figure}

In Fig.~\ref{Ratios} we show the effect of the neutrino mass as 
computed in linear approximation (green, dash-dotted line), one-loop
perturbation theory (black, dotted), and by the TRG method (red,
solid). Blue diamonds are again the results of N-body simulation of
Ref. \cite{Brandbyge:2008js}.The upper (lower) group of lines
corresponds to $M_\nu = 0.3$ ($0.6$) eV, respectively.  From this
figure, we can infer two things. On one hand, the relative change
between massive and massless neutrino models is approximately the same
for the TRG and the one-loop method. This is not surprising,
because the difficulty of the one-loop method to reproduce the N-body
data disappears in the plotted ratios. On the other hand, we see that
both the one-loop and RG method tend to over-estimate the impact of
neutrino masses for small redshift. It should also be noticed that the simulations of Ref.~ \cite{Brandbyge:2008js}, where the neutrinos are modeled by the grid approach, seem to be in closer agreement with us than those in which neutrinos are represented as particles (the blue diamonds in our plots), as could be expected from the conceptual similarity.

In this paper we have presented the derivation of the large scale 
matter power spectrum
in the presence of massive neutrinos using the TRG method which 
resums the  perturbative corrections to the
matter power spectrum to all orders and therefore extends the one-loop
computation. Evaluating numerically the TRG matter power spectrum,
we have seen that nonlinear corrections enhance the suppression of the
power spectrum at small scales due to the neutrino free-streeming. This
enhanced suppression has been observed in N-body simulations
\cite{Brandbyge:2008rv,Brandbyge:2008js} and already appears at one-loop
\cite{Wong:2008ws}, but the TRG is in better agreement with N-body data, as it is manifests from Fig. 2.

Our results indicate that the TRG approach is particularly suited to describe intermediate scales, where linear theory and one-loop perturbation theory are not reliable, and, on the other hand, N-body simulations require very large volumes and are therefore extremely time consuming.  This range of scales will be accurately measured by future galaxy surveys with the aim of pinning down the dark energy equation of state by measuring baryonic acoustic oscillations (BAO), from $k\simeq 0.05 \,h/\mathrm{Mpc}$ to $0.3\, h/\mathrm{Mpc}$. In this context, it has been shown  \cite{CS2, Seo} that non-linearities affect non-trivially the BAO feature by shifting the position of the peaks towards smaller scales with respect to those computed in linear theory. The same effect is witnessed by the wiggles in Figs.~\ref{RatioAbs}, which would not be there if the non-linear corrections were $k$-independent in that range of scales. Neutrino masses, modifiying the non-linear corrections, also have an impact on the BAOs, and this is seen in our Figs.~\ref{Ratios}. 

The approach can be improved along two possible
lines. On one side neutrino perturbations have been treated at the linear 
level; a nonlinear generalization would be of course welcome. On the
other side, the TRG technique might be improved by including
vertex corrections or, in other words, solving the equations
for the correlators including the connected four-point correlation function. 

\acknowledgments
We kindly thank S. Hannestad and J. Brandbyge for sending us the data relative
to 
their N-body simulations with massive neutrinos of Ref. 
\cite{Brandbyge:2008js}, and for comments on the manuscript. Precious comments from Y. Wong are also acknowledged.
This research has been partially supported by ASI contract 
I/016/07/0 "COFIS" and by the European Community's Research Training Networks
under contracts MRTN-CT-2004-50369 and MRTN-CT-2006-035505.
\bibliography{refs}

\begin{thebibliography}{44}
\expandafter\ifx\csname natexlab\endcsname\relax\def\natexlab#1{#1}\fi
\expandafter\ifx\csname bibnamefont\endcsname\relax
  \def\bibnamefont#1{#1}\fi
\expandafter\ifx\csname bibfnamefont\endcsname\relax
  \def\bibfnamefont#1{#1}\fi
\expandafter\ifx\csname citenamefont\endcsname\relax
  \def\citenamefont#1{#1}\fi
\expandafter\ifx\csname url\endcsname\relax
  \def\url#1{\texttt{#1}}\fi
\expandafter\ifx\csname urlprefix\endcsname\relax\def\urlprefix{URL }\fi
\providecommand{\bibinfo}[2]{#2}
\providecommand{\eprint}[2][]{\url{#2}}

\bibitem[{\citenamefont{Schwetz et~al.}(2008)\citenamefont{Schwetz, Tortola,
  and Valle}}]{Schwetz:2008er}
\bibinfo{author}{\bibfnamefont{T.}~\bibnamefont{Schwetz}},
  \bibinfo{author}{\bibfnamefont{M.}~\bibnamefont{Tortola}}, \bibnamefont{and}
  \bibinfo{author}{\bibfnamefont{J.~W.~F.} \bibnamefont{Valle}}
  (\bibinfo{year}{2008}), \eprint{0808.2016}.

\bibitem[{\citenamefont{Mangano et~al.}(2005)}]{Mangano:2005cc}
\bibinfo{author}{\bibfnamefont{G.}~\bibnamefont{Mangano}} \bibnamefont{et~al.},
  \bibinfo{journal}{Nucl. Phys.} \textbf{\bibinfo{volume}{B729}},
  \bibinfo{pages}{221} (\bibinfo{year}{2005}), \eprint{hep-ph/0506164}.

\bibitem[{\citenamefont{Hu et~al.}(1998)\citenamefont{Hu, Eisenstein, and
  Tegmark}}]{Hu:1997mj}
\bibinfo{author}{\bibfnamefont{W.}~\bibnamefont{Hu}},
  \bibinfo{author}{\bibfnamefont{D.~J.} \bibnamefont{Eisenstein}},
  \bibnamefont{and} \bibinfo{author}{\bibfnamefont{M.}~\bibnamefont{Tegmark}},
  \bibinfo{journal}{Phys. Rev. Lett.} \textbf{\bibinfo{volume}{80}},
  \bibinfo{pages}{5255} (\bibinfo{year}{1998}), \eprint{astro-ph/9712057}.

\bibitem[{\citenamefont{Lesgourgues and Pastor}(2006)}]{Lesgourgues:2006nd}
\bibinfo{author}{\bibfnamefont{J.}~\bibnamefont{Lesgourgues}} \bibnamefont{and}
  \bibinfo{author}{\bibfnamefont{S.}~\bibnamefont{Pastor}},
  \bibinfo{journal}{Phys. Rept.} \textbf{\bibinfo{volume}{429}},
  \bibinfo{pages}{307} (\bibinfo{year}{2006}), \eprint{astro-ph/0603494}.

\bibitem[{\citenamefont{Komatsu et~al.}(2008)}]{Komatsu:2008hk}
\bibinfo{author}{\bibfnamefont{E.}~\bibnamefont{Komatsu}} \bibnamefont{et~al.}
  (\bibinfo{collaboration}{WMAP}) (\bibinfo{year}{2008}), \eprint{0803.0547}.

\bibitem[{\citenamefont{Tegmark et~al.}(2006)}]{Tegmark:2006az}
\bibinfo{author}{\bibfnamefont{M.}~\bibnamefont{Tegmark}} \bibnamefont{et~al.}
  (\bibinfo{collaboration}{SDSS}), \bibinfo{journal}{Phys. Rev.}
  \textbf{\bibinfo{volume}{D74}}, \bibinfo{pages}{123507}
  (\bibinfo{year}{2006}), \eprint{astro-ph/0608632}.

\bibitem[{\citenamefont{Song and Knox}(2003)}]{Song:2003gg}
\bibinfo{author}{\bibfnamefont{Y.-S.} \bibnamefont{Song}} \bibnamefont{and}
  \bibinfo{author}{\bibfnamefont{L.}~\bibnamefont{Knox}}
  (\bibinfo{year}{2003}), \eprint{astro-ph/0312175}.

\bibitem[{\citenamefont{Hannestad et~al.}(2006)\citenamefont{Hannestad, Tu, and
  Wong}}]{Hannestad:2006as}
\bibinfo{author}{\bibfnamefont{S.}~\bibnamefont{Hannestad}},
  \bibinfo{author}{\bibfnamefont{H.}~\bibnamefont{Tu}}, \bibnamefont{and}
  \bibinfo{author}{\bibfnamefont{Y.~Y.~Y.} \bibnamefont{Wong}},
  \bibinfo{journal}{JCAP} \textbf{\bibinfo{volume}{0606}}, \bibinfo{pages}{025}
  (\bibinfo{year}{2006}), \eprint{astro-ph/0603019}.

\bibitem[{\citenamefont{Kitching et~al.}(2008)\citenamefont{Kitching, Heavens,
  Verde, Serra, and Melchiorri}}]{Kitching:2008dp}
\bibinfo{author}{\bibfnamefont{T.~D.} \bibnamefont{Kitching}},
  \bibinfo{author}{\bibfnamefont{A.~F.} \bibnamefont{Heavens}},
  \bibinfo{author}{\bibfnamefont{L.}~\bibnamefont{Verde}},
  \bibinfo{author}{\bibfnamefont{P.}~\bibnamefont{Serra}}, \bibnamefont{and}
  \bibinfo{author}{\bibfnamefont{A.}~\bibnamefont{Melchiorri}},
  \bibinfo{journal}{Phys. Rev.} \textbf{\bibinfo{volume}{D77}},
  \bibinfo{pages}{103008} (\bibinfo{year}{2008}), \eprint{0801.4565}.

\bibitem[{\citenamefont{Lesgourgues et~al.}(2006)\citenamefont{Lesgourgues,
  Perotto, Pastor, and Piat}}]{Lesgourgues:2005yv}
\bibinfo{author}{\bibfnamefont{J.}~\bibnamefont{Lesgourgues}},
  \bibinfo{author}{\bibfnamefont{L.}~\bibnamefont{Perotto}},
  \bibinfo{author}{\bibfnamefont{S.}~\bibnamefont{Pastor}}, \bibnamefont{and}
  \bibinfo{author}{\bibfnamefont{M.}~\bibnamefont{Piat}},
  \bibinfo{journal}{Phys. Rev.} \textbf{\bibinfo{volume}{D73}},
  \bibinfo{pages}{045021} (\bibinfo{year}{2006}), \eprint{astro-ph/0511735}.

\bibitem[{\citenamefont{Perotto et~al.}(2006)\citenamefont{Perotto,
  Lesgourgues, Hannestad, Tu, and Wong}}]{Perotto:2006rj}
\bibinfo{author}{\bibfnamefont{L.}~\bibnamefont{Perotto}},
  \bibinfo{author}{\bibfnamefont{J.}~\bibnamefont{Lesgourgues}},
  \bibinfo{author}{\bibfnamefont{S.}~\bibnamefont{Hannestad}},
  \bibinfo{author}{\bibfnamefont{H.}~\bibnamefont{Tu}}, \bibnamefont{and}
  \bibinfo{author}{\bibfnamefont{Y.~Y.~Y.} \bibnamefont{Wong}},
  \bibinfo{journal}{JCAP} \textbf{\bibinfo{volume}{0610}}, \bibinfo{pages}{013}
  (\bibinfo{year}{2006}), \eprint{astro-ph/0606227}.

\bibitem[{\citenamefont{Wang et~al.}(2005)\citenamefont{Wang, Haiman, Hu,
  Khoury, and May}}]{Wang:2005vr}
\bibinfo{author}{\bibfnamefont{S.}~\bibnamefont{Wang}},
  \bibinfo{author}{\bibfnamefont{Z.}~\bibnamefont{Haiman}},
  \bibinfo{author}{\bibfnamefont{W.}~\bibnamefont{Hu}},
  \bibinfo{author}{\bibfnamefont{J.}~\bibnamefont{Khoury}}, \bibnamefont{and}
  \bibinfo{author}{\bibfnamefont{M.}~\bibnamefont{May}},
  \bibinfo{journal}{Phys. Rev. Lett.} \textbf{\bibinfo{volume}{95}},
  \bibinfo{pages}{011302} (\bibinfo{year}{2005}), \eprint{astro-ph/0505390}.

\bibitem[{\citenamefont{Wyithe and Loeb}(2008)}]{Wyithe:2008mv}
\bibinfo{author}{\bibfnamefont{S.}~\bibnamefont{Wyithe}} \bibnamefont{and}
  \bibinfo{author}{\bibfnamefont{A.}~\bibnamefont{Loeb}}
  (\bibinfo{year}{2008}), \eprint{0808.2323}.

\bibitem[{\citenamefont{Pritchard and Pierpaoli}(2008)}]{Pritchard:2008wy}
\bibinfo{author}{\bibfnamefont{J.~R.} \bibnamefont{Pritchard}}
  \bibnamefont{and} \bibinfo{author}{\bibfnamefont{E.}~\bibnamefont{Pierpaoli}}
  (\bibinfo{year}{2008}), \eprint{0805.1920}.

\bibitem[{\citenamefont{Hannestad and Wong}(2007)}]{Hannestad:2007cp}
\bibinfo{author}{\bibfnamefont{S.}~\bibnamefont{Hannestad}} \bibnamefont{and}
  \bibinfo{author}{\bibfnamefont{Y.~Y.~Y.} \bibnamefont{Wong}},
  \bibinfo{journal}{JCAP} \textbf{\bibinfo{volume}{0707}}, \bibinfo{pages}{004}
  (\bibinfo{year}{2007}), \eprint{astro-ph/0703031}.

\bibitem[{\citenamefont{Ichikawa and Takahashi}(2008)}]{Ichikawa:2005hi}
\bibinfo{author}{\bibfnamefont{K.}~\bibnamefont{Ichikawa}} \bibnamefont{and}
  \bibinfo{author}{\bibfnamefont{T.}~\bibnamefont{Takahashi}},
  \bibinfo{journal}{JCAP} \textbf{\bibinfo{volume}{0802}}, \bibinfo{pages}{017}
  (\bibinfo{year}{2008}), \eprint{astro-ph/0510849}.

\bibitem[{\citenamefont{Lesgourgues et~al.}(2008)\citenamefont{Lesgourgues,
  Valkenburg, and Gaztanaga}}]{Lesgourgues:2007ix}
\bibinfo{author}{\bibfnamefont{J.}~\bibnamefont{Lesgourgues}},
  \bibinfo{author}{\bibfnamefont{W.}~\bibnamefont{Valkenburg}},
  \bibnamefont{and}
  \bibinfo{author}{\bibfnamefont{E.}~\bibnamefont{Gaztanaga}},
  \bibinfo{journal}{Phys. Rev.} \textbf{\bibinfo{volume}{D77}},
  \bibinfo{pages}{063505} (\bibinfo{year}{2008}), \eprint{0710.5525}.

\bibitem[{\citenamefont{Gratton et~al.}(2008)\citenamefont{Gratton, Lewis, and
  Efstathiou}}]{Gratton:2007tb}
\bibinfo{author}{\bibfnamefont{S.}~\bibnamefont{Gratton}},
  \bibinfo{author}{\bibfnamefont{A.}~\bibnamefont{Lewis}}, \bibnamefont{and}
  \bibinfo{author}{\bibfnamefont{G.}~\bibnamefont{Efstathiou}},
  \bibinfo{journal}{Phys. Rev.} \textbf{\bibinfo{volume}{D77}},
  \bibinfo{pages}{083507} (\bibinfo{year}{2008}), \eprint{0705.3100}.

\bibitem[{\citenamefont{Tereno et~al.}(2008)}]{Tereno:2008mm}
\bibinfo{author}{\bibfnamefont{I.}~\bibnamefont{Tereno}} \bibnamefont{et~al.}
  (\bibinfo{year}{2008}), \eprint{0810.0555}.

\bibitem[{\citenamefont{Ichiki et~al.}(2008)\citenamefont{Ichiki, Takada, and
  Takahashi}}]{Ichiki:2008ye}
\bibinfo{author}{\bibfnamefont{K.}~\bibnamefont{Ichiki}},
  \bibinfo{author}{\bibfnamefont{M.}~\bibnamefont{Takada}}, \bibnamefont{and}
  \bibinfo{author}{\bibfnamefont{T.}~\bibnamefont{Takahashi}}
  (\bibinfo{year}{2008}), \eprint{0810.4921}.

\bibitem[{\citenamefont{Abazajian et~al.}(2005)\citenamefont{Abazajian,
  Switzer, Dodelson, Heitmann, and Habib}}]{Abazajian:2004zh}
\bibinfo{author}{\bibfnamefont{K.}~\bibnamefont{Abazajian}},
  \bibinfo{author}{\bibfnamefont{E.~R.} \bibnamefont{Switzer}},
  \bibinfo{author}{\bibfnamefont{S.}~\bibnamefont{Dodelson}},
  \bibinfo{author}{\bibfnamefont{K.}~\bibnamefont{Heitmann}}, \bibnamefont{and}
  \bibinfo{author}{\bibfnamefont{S.}~\bibnamefont{Habib}},
  \bibinfo{journal}{Phys. Rev.} \textbf{\bibinfo{volume}{D71}},
  \bibinfo{pages}{043507} (\bibinfo{year}{2005}), \eprint{astro-ph/0411552}.

\bibitem[{\citenamefont{Dore and Aghanim}(2009)}]{Dore}
\bibinfo{author}{\bibfnamefont{O.}~\bibnamefont{Dore}} \bibnamefont{and}
  \bibinfo{author}{\bibfnamefont{N.}~\bibnamefont{Aghanim}},
  \bibinfo{journal}{in preparation}  (\bibinfo{year}{2009}).

\bibitem[{\citenamefont{Klypin et~al.}(1993)\citenamefont{Klypin, Holtzman,
  Primack, and Regos}}]{Klypin}
\bibinfo{author}{\bibfnamefont{A.}~\bibnamefont{Klypin}},
  \bibinfo{author}{\bibfnamefont{J.}~\bibnamefont{Holtzman}},
  \bibinfo{author}{\bibfnamefont{J.}~\bibnamefont{Primack}}, \bibnamefont{and}
  \bibinfo{author}{\bibfnamefont{E.}~\bibnamefont{Regos}},
  \bibinfo{journal}{Astrophys. J.} \textbf{\bibinfo{volume}{416}},
  \bibinfo{pages}{1} (\bibinfo{year}{1993}), \eprint{astro-ph/9305011}.

\bibitem[{\citenamefont{Primack et~al.}(1995)\citenamefont{Primack, Holtzman,
  Klypin, and Caldwell}}]{Primack}
\bibinfo{author}{\bibfnamefont{J.~R.} \bibnamefont{Primack}},
  \bibinfo{author}{\bibfnamefont{J.}~\bibnamefont{Holtzman}},
  \bibinfo{author}{\bibfnamefont{A.}~\bibnamefont{Klypin}}, \bibnamefont{and}
  \bibinfo{author}{\bibfnamefont{D.~O.} \bibnamefont{Caldwell}},
  \bibinfo{journal}{Phys. Rev. Lett.} \textbf{\bibinfo{volume}{74}},
  \bibinfo{pages}{2160} (\bibinfo{year}{1995}), \eprint{astro-ph/9411020}.

\bibitem[{\citenamefont{Brandbyge et~al.}(2008)\citenamefont{Brandbyge,
  Hannestad, Haugboelle, and Thomsen}}]{Brandbyge:2008rv}
\bibinfo{author}{\bibfnamefont{J.}~\bibnamefont{Brandbyge}},
  \bibinfo{author}{\bibfnamefont{S.}~\bibnamefont{Hannestad}},
  \bibinfo{author}{\bibfnamefont{T.}~\bibnamefont{Haugboelle}},
  \bibnamefont{and} \bibinfo{author}{\bibfnamefont{B.}~\bibnamefont{Thomsen}},
  \bibinfo{journal}{JCAP} \textbf{\bibinfo{volume}{0808}}, \bibinfo{pages}{020}
  (\bibinfo{year}{2008}), \eprint{0802.3700}.

\bibitem[{\citenamefont{Brandbyge and Hannestad}(2008)}]{Brandbyge:2008js}
\bibinfo{author}{\bibfnamefont{J.}~\bibnamefont{Brandbyge}} \bibnamefont{and}
  \bibinfo{author}{\bibfnamefont{S.}~\bibnamefont{Hannestad}}
  (\bibinfo{year}{2008}), \eprint{0812.3149}.

\bibitem[{\citenamefont{Bernardeau et~al.}(2002)\citenamefont{Bernardeau,
  Colombi, Gaztanaga, and Scoccimarro}}]{PTreview}
\bibinfo{author}{\bibfnamefont{F.}~\bibnamefont{Bernardeau}},
  \bibinfo{author}{\bibfnamefont{S.}~\bibnamefont{Colombi}},
  \bibinfo{author}{\bibfnamefont{E.}~\bibnamefont{Gaztanaga}},
  \bibnamefont{and}
  \bibinfo{author}{\bibfnamefont{R.}~\bibnamefont{Scoccimarro}},
  \bibinfo{journal}{Phys. Rept.} \textbf{\bibinfo{volume}{367}},
  \bibinfo{pages}{1} (\bibinfo{year}{2002}), \eprint{astro-ph/0112551}.

\bibitem[{\citenamefont{Jeong and Komatsu}(2006)}]{JK06}
\bibinfo{author}{\bibfnamefont{D.}~\bibnamefont{Jeong}} \bibnamefont{and}
  \bibinfo{author}{\bibfnamefont{E.}~\bibnamefont{Komatsu}},
  \bibinfo{journal}{Astrophys. J.} \textbf{\bibinfo{volume}{651}},
  \bibinfo{pages}{619} (\bibinfo{year}{2006}), \eprint{astro-ph/0604075}.

\bibitem[{\citenamefont{Jeong and Komatsu}(2008)}]{JK08}
\bibinfo{author}{\bibfnamefont{D.}~\bibnamefont{Jeong}} \bibnamefont{and}
  \bibinfo{author}{\bibfnamefont{E.}~\bibnamefont{Komatsu}}
  (\bibinfo{year}{2008}), \eprint{0805.2632}.

\bibitem[{\citenamefont{Crocce and Scoccimarro}(2006)}]{CS1}
\bibinfo{author}{\bibfnamefont{M.}~\bibnamefont{Crocce}} \bibnamefont{and}
  \bibinfo{author}{\bibfnamefont{R.}~\bibnamefont{Scoccimarro}},
  \bibinfo{journal}{Phys. Rev.} \textbf{\bibinfo{volume}{D73}},
  \bibinfo{pages}{063519} (\bibinfo{year}{2006}), \eprint{astro-ph/0509418}.

\bibitem[{\citenamefont{Crocce and Scoccimarro}(2008)}]{CS2}
\bibinfo{author}{\bibfnamefont{M.}~\bibnamefont{Crocce}} \bibnamefont{and}
  \bibinfo{author}{\bibfnamefont{R.}~\bibnamefont{Scoccimarro}},
  \bibinfo{journal}{Phys. Rev.} \textbf{\bibinfo{volume}{D77}},
  \bibinfo{pages}{023533} (\bibinfo{year}{2008}), \eprint{0704.2783}.

\bibitem[{\citenamefont{Taruya and Hiramatsu}(2007)}]{Taruya2007}
\bibinfo{author}{\bibfnamefont{A.}~\bibnamefont{Taruya}} \bibnamefont{and}
  \bibinfo{author}{\bibfnamefont{T.}~\bibnamefont{Hiramatsu}}
  (\bibinfo{year}{2007}), \eprint{0708.1367}.

\bibitem[{\citenamefont{McDonald}(2007)}]{McD06}
\bibinfo{author}{\bibfnamefont{P.}~\bibnamefont{McDonald}},
  \bibinfo{journal}{Phys. Rev.} \textbf{\bibinfo{volume}{D75}},
  \bibinfo{pages}{043514} (\bibinfo{year}{2007}), \eprint{astro-ph/0606028}.

\bibitem[{\citenamefont{Matarrese and Pietroni}(2008)}]{MP07a}
\bibinfo{author}{\bibfnamefont{S.}~\bibnamefont{Matarrese}} \bibnamefont{and}
  \bibinfo{author}{\bibfnamefont{M.}~\bibnamefont{Pietroni}},
  \bibinfo{journal}{Mod. Phys. Lett.} \textbf{\bibinfo{volume}{A23}},
  \bibinfo{pages}{25} (\bibinfo{year}{2008}), \eprint{astro-ph/0702653}.

\bibitem[{\citenamefont{Matarrese and Pietroni}(2007)}]{MP07b}
\bibinfo{author}{\bibfnamefont{S.}~\bibnamefont{Matarrese}} \bibnamefont{and}
  \bibinfo{author}{\bibfnamefont{M.}~\bibnamefont{Pietroni}},
  \bibinfo{journal}{JCAP} \textbf{\bibinfo{volume}{0706}}, \bibinfo{pages}{026}
  (\bibinfo{year}{2007}), \eprint{astro-ph/0703563}.

\bibitem[{\citenamefont{Izumi and Soda}(2007)}]{Izumi07}
\bibinfo{author}{\bibfnamefont{K.}~\bibnamefont{Izumi}} \bibnamefont{and}
  \bibinfo{author}{\bibfnamefont{J.}~\bibnamefont{Soda}},
  \bibinfo{journal}{Phys. Rev.} \textbf{\bibinfo{volume}{D76}},
  \bibinfo{pages}{083517} (\bibinfo{year}{2007}), \eprint{0706.1604}.

\bibitem[{\citenamefont{Matsubara}(2008{\natexlab{a}})}]{Matsubara07}
\bibinfo{author}{\bibfnamefont{T.}~\bibnamefont{Matsubara}},
  \bibinfo{journal}{Phys. Rev.} \textbf{\bibinfo{volume}{D77}},
  \bibinfo{pages}{063530} (\bibinfo{year}{2008}{\natexlab{a}}),
  \eprint{0711.2521}.

\bibitem[{\citenamefont{Valageas}(2007)}]{ValBisp}
\bibinfo{author}{\bibfnamefont{P.}~\bibnamefont{Valageas}}
  (\bibinfo{year}{2007}), \eprint{0711.3407}.

\bibitem[{\citenamefont{Matsubara}(2008{\natexlab{b}})}]{Matsubara:2008wx}
\bibinfo{author}{\bibfnamefont{T.}~\bibnamefont{Matsubara}},
  \bibinfo{journal}{Phys. Rev.} \textbf{\bibinfo{volume}{D78}},
  \bibinfo{pages}{083519} (\bibinfo{year}{2008}{\natexlab{b}}),
  \eprint{0807.1733}.

\bibitem[{\citenamefont{Pietroni}(2008)}]{P08a}
\bibinfo{author}{\bibfnamefont{M.}~\bibnamefont{Pietroni}},
  \bibinfo{journal}{JCAP} \textbf{\bibinfo{volume}{0810}}, \bibinfo{pages}{036}
  (\bibinfo{year}{2008}), \eprint{0806.0971}.

\bibitem[{\citenamefont{Saito et~al.}(2008)\citenamefont{Saito, Takada, and
  Taruya}}]{Saito:2008bp}
\bibinfo{author}{\bibfnamefont{S.}~\bibnamefont{Saito}},
  \bibinfo{author}{\bibfnamefont{M.}~\bibnamefont{Takada}}, \bibnamefont{and}
  \bibinfo{author}{\bibfnamefont{A.}~\bibnamefont{Taruya}},
  \bibinfo{journal}{Phys. Rev. Lett.} \textbf{\bibinfo{volume}{100}},
  \bibinfo{pages}{191301} (\bibinfo{year}{2008}), \eprint{0801.0607}.

\bibitem[{\citenamefont{Wong}(2008)}]{Wong:2008ws}
\bibinfo{author}{\bibfnamefont{Y.~Y.~Y.} \bibnamefont{Wong}},
  \bibinfo{journal}{JCAP} \textbf{\bibinfo{volume}{0810}}, \bibinfo{pages}{035}
  (\bibinfo{year}{2008}), \eprint{0809.0693}.

\bibitem[{\citenamefont{Lewis et~al.}(2000)\citenamefont{Lewis, Challinor, and
  Lasenby}}]{Lewis:1999bs}
\bibinfo{author}{\bibfnamefont{A.}~\bibnamefont{Lewis}},
  \bibinfo{author}{\bibfnamefont{A.}~\bibnamefont{Challinor}},
  \bibnamefont{and} \bibinfo{author}{\bibfnamefont{A.}~\bibnamefont{Lasenby}},
  \bibinfo{journal}{Astrophys. J.} \textbf{\bibinfo{volume}{538}},
  \bibinfo{pages}{473} (\bibinfo{year}{2000}), \eprint{astro-ph/9911177}.

\bibitem[{\citenamefont{Seo et~al.}(2008)\citenamefont{Seo, Siegel, Eisenstein,
  and White}}]{Seo}
\bibinfo{author}{\bibfnamefont{H.-J.} \bibnamefont{Seo}},
  \bibinfo{author}{\bibfnamefont{E.~R.} \bibnamefont{Siegel}},
  \bibinfo{author}{\bibfnamefont{D.~J.} \bibnamefont{Eisenstein}},
  \bibnamefont{and} \bibinfo{author}{\bibfnamefont{M.}~\bibnamefont{White}}
  (\bibinfo{year}{2008}), \eprint{0805.0117}.

\end{thebibliography}
\end{document}